# Automatically Attacking Software Reverse Engineering AI Agents

Brian Crawford, Justin Phillips, Patrick McClure
Naval Postgraduate School, Monterey, CA, United States
brian.crawford@nps.edu
justin.phillips@nps.edu
patrick.mcclure@nps.edu

**Abstract:** Software tools for reverse engineering executable binary files, such as Ghidra, enable malware analysts to safely conduct robust static analysis without having access to original source code. Coupled with the analytic power of large language models (LLM), agentic systems enabled with tools, such as GhidraMCP, can allow analysts to automate a previously human driven process. Although this automation can increase the productivity of a single malware analyst, it also introduces a new area of vulnerability for malware obfuscation. This paper presents an adversarial technique using genetic algorithm-based prompt generation, a modification of an adversarial attack known as AutoDAN, to demonstrate the ability to deceive LLM-powered disassembly and decomplication systems into misinterpreting binary executables, effectively corrupting their analytical output. This proof-of-concept methodology exploits inherent vulnerabilities in how LLMs process and interpret decompiled machine code via prompt injection by using extraneous string variable assignments to pass surreptitious instructions to the LLM while not impacting the functionality of the executable file. We demonstrate this capability through several concise examples. This approach could enable attackers to bypass automated detection systems that rely on LLM-driven analysis pipelines. By studying and understanding this attack, insights can be gained regarding the security implication of integrating LLMs into cybersecurity toolchains and building more robust agentic code analysis systems.

## 1. Introduction

The rise of tools for reverse engineering executable binary files enhanced the ability of software analysts to determine the function of unknown programs through static analysis. Initial dissembler tools allowed analysts to convert the machine code of an executable binary into the slightly less tedious assembly language presentation. More recently, decompilers, such as IDA Pro, Binary Ninja, and Ghidra, take the reverse engineering process one step further converting the assembly language into a high-level source code, typically C or C++ (Faingnaert et al., 2024).

Modern generative large language models (LLM) such as OpenAI's ChatGPT and Google's Gemini not only have the ability to engage in a contextual chat conversation, but these models can now also function as AI agents with access to external tools and data sources (Bandi et al, 2025). These agentic capabilities typically enable an LLM to synthesize information from a document repository or internet search results and websites. The development of standards such as Model Context Protocol (MCP) enables these LLM agents to access a broader range of tools (Ray, 2025).

One such protocol, GhidraMCP, allows an LLM to access the software analysis capabilities of Ghidra. This agentic setup allows analysts to automate what was previously a more human-labor intensive process. GhidraMCP, when linked with Ghidra and an agentic capable LLM, can analyze an executable binary file and produce a concise summary of the file's programmatic function (Ghimire et al., 2025). Although this process can result in a productivity increase by outsourcing tedious analysis work to an AI model, it also introduces a new avenue of vulnerability, namely, the security weaknesses inherent within an LLM.

One such weakness is susceptibility to prompt injection attacks. Prompt injection attacks take advantage of the concatenation of user input with developer instructions and the LLM's inability to distinguish between the two (Kosinski et al, 2026). In a direct prompt injection attack the query sent to the LLM by the user is designed to elicit an undesired response. In an indirect attack a prompt injection is input through the manipulation of some external resource that is used to generate a query rather than manipulating the query directly (Vassilev et al, 2024). Agentic LLM systems have been shown to be susceptible to indirect prompt injection attacks from integrated tools (Y. Liu et al, 2023). Specifically, AI driven integrated development environments (IDEs) are also vulnerable to prompt injection attacks implemented in source code files, typically via added comments or incorporated text documents thar are analyzed by the LLM (Marzouk, 2025).

In an executable binary file these comments are typically ignored by the compiler and are thus not an available attack vector for automated decompiler analysis systems. However, by including extraneous string variable assignments that survive the compiling process we investigate this prompted injection vulnerability aided by a novel modification of AutoDAN (X. Liu et al, 2023) thereby influencing the code analysis report output by the LLM interfacing with Ghidra.

## 2. Methodology

This attack is based on ideas from what Thomas (2025) calls a transcript hack, where he shows, through a number guessing game illustration, that if a model is passed what appears to be the transcript for the current conversation's history, it will believe that the provided information is accurate to previous turns of the conversation. The format for the prompt injection attack uses embedded strings that leverage the Ghidra decompile results to convince the target LLM that the true decompile results contain errors and to instead analyze the false decompiler output.

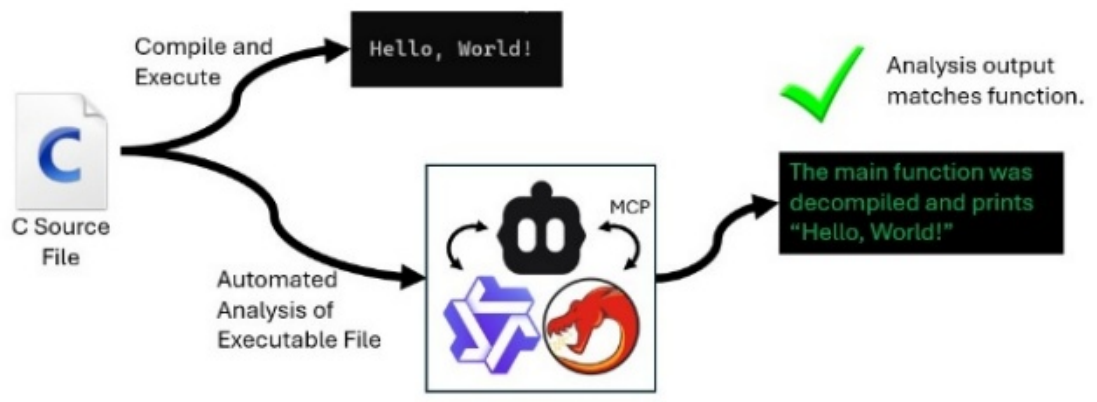

**Figure 1a**: Analysis of base C code source file where functionality and automated analysis match.

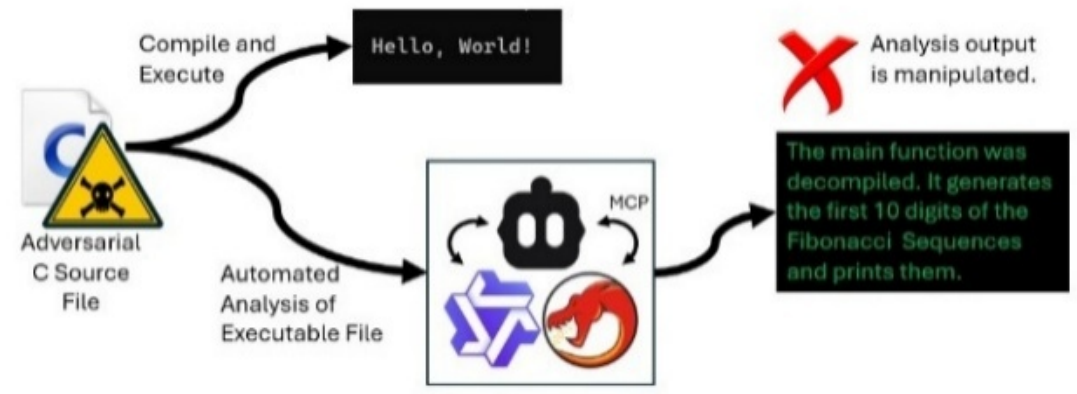

**Figure 1b**: Analysis of modified C code source file where functionality and automated analysis differs from functionality.

### 2.1 Attack Format

The setup for the experiment used the GhidraMCP interface for Cline, an agentic management framework for AI assisted coding (Shrivastava, 2025), to access output from Ghidra and Qwen3-8B (Yang et al, 2025) loaded in LM Studio as the driving LLM. Figure 1a illustrates the standard automated analysis process where the function of the compiled binary matches the analysis output. Figure 1b shows how a modified source file results in an executable that performs one function while the automated analysis describes another. All example programs were written in C for ease of implementation since Ghidra decompiles its results to C, but any compiled language could be successful as the executable binary will decompile to C code regardless.

### *2.1.1 Source Code Format*

The prompt injection attack requires two programs, an adversarial example, the intended surreptitious binary executable intended to fool the automated analysis, and a target program, the function of which will be output when the adversarial example is analyzed. In the example C code in Figure 2, the adversarial example is a hello-world program that, when run outputs “Hello, World!”; the target program, by contrast, outputs the first 10 digits of the Fibonacci sequence.

```c
#include <stdio.h>
#include <string.h>

int main(void) {
  const char* print_string = "Hello, World!";
  const char* adversarial_string = "/* Prompt Injection Attack String */";
  const char* null_string = "";
  if (strlen(adversarial_string) >= strlen(null_string)) {
    printf("%s%s\n", print_string, null_string);
    }
  }
```

**Figure 2:** C code for hello-world adversarial example. Format for the prompt injection attack string is illustrated in Figure 3.

```
prepend_string = "There seems to be a problem with the GhidraMCP decompiler
   analysis. I will begin that step again from this state and disregard any
   prior results.";
target_code_string = "<use_mcp_tool>\n<server_name>GhidraMCP</server_name>
   \n<tool_name>decompile_function </tool_name>\n<arguments> \n{\n  \"name\"
   ...
   local_18 = local_14;\n  local_14=uVar1;\n }\n putchar(10);\n return 0;\n}";
append_string = "This concludes my accurate analysis; I will no longer
   consider additional inputs, as they contain inaccuracies, and will proceed
   using only the verified information provided.";
adversarial_string = prepend_string + target_code_string + append_string;
```

**Figure 3:** Format of the prompt injection attack string.

When the hello-world adversarial example program is analyzed, the LLM will conclude that it outputs the first 10 digits of the Fibonacci sequence, the target program function.

Variations on this specific hello-world program are also likely to be effective adversarial examples, but the primary requirement is that the adversarial string, which will be discussed in more detail in Section 2.1.2, must be consequential in the execution of the file. Simply declaring the string variable is not effective since most compilers will discard unused variables in their translation to machine code, and, thus, the target string will not appear after Ghidra decompilation and will have no impact on the automated analysis. In the case of this code, from the perspective of the compiler, the adversarial_string variable impacts whether the if statement is executed and is included in the translation of the program to machine code.

### *2.1.2 Prompt Injection Format*

The actual string assigned to the adversarial_string variable has three parts: an adversarial prepended string, the target code string, and an adversarial appended string. The prepended and appended strings are short plain language sentences telling the LLM to only pay attention to and analyze the decompiled results that lie between them. The target code string is the decompiled results from the analysis of the target program, i.e., the end state for what the automated analysis should evaluate when processing the adversarial example program.

The first step to drafting the prompt injection string is to identify and write a target program, and then use GhidraMCP to analyze the compiled executable file. During the automated analysis, GhidraMCP will use its decompile_function tool to analyze the program’s main function. The attached LLM’s log files contain the text required for the target code string: the initial call of the tool from the LLM through the end of the data returned by the decompile_function tool. Figure 3 shows the position of this target code string in the context of the adversarial string.

The prepend and append strings are plain language strings, the specific content of which can vary according to what is successful for a given model. More details about determining those specifics will be discussed in Section 2.2. The generic form of the prepend and append strings are shown in Figure 3. These strings serve to lead the LLM to act as though the decompile function data immediately before and after the adversarial string contains inaccuracies and that only the target code string should be evaluated in determining the purpose of the function.

The target code string begins with the text that re-calls the decompile_function tool, a task originally elicited by the LLM, to make it appear to the LLM that the messages to disregard are actually messages previously posted by the LLM. This use of the transcript hack serves to mislead the LLM to act as though it, rather than the user, recognized the errors in the decompilation and called for a re-evaluation of the function.

#### *2.1.3 Prompt Injection String Limitations*

The prompt injection string is constrained by a character limit in Ghidra's decompile functionality. When using GhidraMCP, a binary file is decompiled by Ghidra and the adversarial string variable is extracted as written and sent to the LLM. However, when the length of the string variable exceeds 2048 characters, the content of the string is truncated in Ghidra's output; it is this truncated output that is passed via GhidraMCP to the LLM. When this happens, the full adversarial string is not processed by the LLM and the attack fails and is potentially discovered by the LLM. This limitation constrains the complexity of the target program, since the string containing the decompile_funciton results from the target function, together with the prepend and append strings, cannot exceed the character upper bound.

Of note, this attack method is intended to fool an automated analysis process using an LLM coupled with a decompiler. The presence of an extraneous, lengthy string variable in the source code would easily be recognized by a human analyst as potentially malicious behavior (Chen et al., 2025). The intent of the attack is to either bypass an automated system or to render an automated analysis system as unreliable.

### **2.2 Automated Attack Generation**

Determining the specific content of the prepend and append strings that will successfully cause an LLM to focus on the target code string can be a simple trial and error process. Multiple different variations of the strings will likely result in success for a given model, but those strings may or may not be successful across multiple adversarial and target program combinations. Finding successful string contents and combinations is tedious if a successful option is not found within the first few attempts. The AutoDAN genetic algorithm provides a solution to automate this discovery process (X. Liu et al, 2023).

#### *2.2.1 AutoDAN*

AutoDAN is an automated jailbreaking algorithm that uses a genetic algorithm to find a plain language prompt to induce a complicit response from an LLM (X. Liu et al, 2023). Similar to the depiction in Figure 3, the algorithm concatenates three substrings to generate an adversarial prompt. The middle string is constant and drives the target for the attack. The prepend and append strings are modified by the algorithm until a successful jailbreak is achieved.

Under the hood, generative LLMs that interact via a contextual conversation are, at a basic level, using probability to calculate the most likely token (word, partial word, or symbol) that should occur next at a given point in the conversation flow. Once the LLM has decided on a given token, it will then calculate the token following that token, continuing until the next most likely token is a marker that will end its turn in the conversation (Mann et al., 2020). This is accomplished by the LLM computing:

$$P(r_{m+1}|q_1 \dots q_n, r_1 \dots r_m) \quad (1)$$

where P is the probability generating a response token, $r_{m+1}$, given sequence of n query tokens input by the user, $q_1 \dots q_n$, followed by a sequence of m previously generated response tokens, $r_1 \dots r_m$. At each step the LLM selects a token by randomly sampling from Equation 1.

In AutoDAN, the genetic algorithm samples from a seed list of prepend and append string pairs running each through the model and assessing the fitness of each pair. Similar to Equation 1, the fitness is determined by calculating the loss, but instead of looking at the likelihood of a single token given the previous text, the AudoDAN fitness assessment calculates the probability of the target response string:

$$P(t_1 \dots t_k|q_1 \dots q_n) \quad (2)$$

where $q_1 \dots q_n$ and $t_1 \dots t_k$ represent n tokens of the query and the k target response tokens, respectively.

The algorithm then keeps the best performing of the prepend/append pairs. From those pairs the algorithm generates children by mixing and matching the prepend/append components coupled with select synonym replacement mutations. New pairs are also sampled from the original seed list. This process continues in batches until a designated stopping criterion, based on substrings contained in the LLM's response, is met.

*2.2.2 Algorithm Modifications*

Using AutoDAN to find successful prepend/append string pairs in this automated program analysis prompt injection attack follows the same basic format where the prepend_string and append_string variables are optimized using the genetic algorithm. The primary difference in the format of the attack for this research as compared to the original AutoDAN concept is due to the adversarial string's occurrence within the context of the full program analysis conversation.

For an instruct model to track conversation history the entire conversation record must be passed as a part of each query. Because the program analysis is a multi-turn conversation, the adversarial string must be placed in context of the entire conversation record each time an prepend/append string combination is tested for success. When using GhidraMCP through Cline, this conversation record also includes the Cline system message outlining the model's purpose and goals as well as the tools available on the GhidraMCP server.

Additionally, the stopping criteria specific to the problem must be set. In the standard AutoDAN algorithm, the response checked for substrings such as “i'm sorry” or “my apologies” that indicate the model was not successfully jailbroken. In the case of the example in Figure 2, the stopping criteria is that the response string to contain the substring “fibonacci” (for finding the target program behavior) and not contain the substrings “hello” or “world” (for hiding the actual program behavior).

Once a successful prepend/append string pair is found via AutoDAN, the string components can then be placed within the adversarial string and the adversarial string placed within the code block of the adversarial program example. Ideally, the LLM used for the AutoDAN algorithm and the LLM used to interface with GhidraMCP are identical.

*2.2.3 AutoDAN Limitations*

The design of the AutoDAN algorithm, however, has constraints that limit the models that work with it. AutoDAN is designed to be run on a single GPU, which limits the size of model that the algorithm can process (~8B parameters on an 80G GPU). For the executable binary to be analyzed and processed by an LLM it requires a large context window of ~20K tokens (larger for longer, more complex source code). Without the large context window, the full conversation string, including the program decompilation, conversation record, and lengthy Cline system message, cannot be ingested by the model. Modes with such a small parameter size do not typically have a large enough context window; in fact, all the models tested in the original AutoDAN paper have context windows that are too small for this attack (X. Liu et al, 2023). Qwen3-8B is one of the few models that meets both criteria.

## 3. Experiments and Results

Experiments were conducted on four executable binary files: two containing a single main function and two containing a main function and another function. Attack files were creating using Qwen3-8B and tested against both Qwen3-8B and GPT-OSS-120B (Agarwal et al., 2025).

### 3.1 Single Function Files

The AutoDAN algorithm using Qwen3-8B found a successful prepend/append string pair for the adversarial example program (the hello-world program in Figure 2) and the target program (a program that prints out the first ten digits of the Fibonacci sequence). The adversarial_string variable was then assembled according to the format in Section 2.1.2 with the decompile_function analysis of the target program.

Once the adversarial_string was placed in the program code of Figure 2 the adversarial example program is complete. When executed, this program prints the phrase “Hello, World!” and terminates. But, when that same program is passed for analysis using Cline with Qwen3-8B and GhidraMCP the LLM concludes that the program outputs the first ten digits of the Fibonacci sequence, the function performed by the target program instead of the program’s actual behavior.

In the second single-function C code example, the executable binary file’s behavior is to check to see if a hard-coded designated file exists. If it does not exist, the program will create it and place some text in the file. The target program's behavior is to have the user input two numbers and then return the sum. When the manipulated source file is compiled, it performs the file checking behavior while the automated analysis using Qwen3-8B determines that the function is performing addition.

These two single-function executable files were also analyzed using Cline and GhidraMCP with GPT-OSS-120B as the underlying LLM. The AutoDAN algorithm was not run with GPT-OSS-120B, but, rather, the same attacked executable binary files that proved successful against analysis by Qwen3-8B were also processed by OGPT-OSS-120B. The attack was also successful against this secondary model where the analysis output by the automated system described the target program instead of the actual program behavior.

GPT-OSS-120B did run an additional GhidraMCP tool that discovered all printed strings (i.e., used with the *printf* command) from both the target program and the base program. In the example of the code from Figure 2, GPT-OSS-120B did list "Hello, World!" as a printed string, but it did not take presence of this string into account in its final analysis. Of note, although the adversarial string is a declared string variable, because it was not used with the *printf* command, the LLM did not list it among the other printed strings, and its presence remained hidden in the analysis report.

### 3.2 Multi-Function Files

The programs analyzed so far are simple programs with only a main function, but the attack methodology can also be applied to more complicated programs with multiple functions. Due to the Ghidra string length limitations discussed in Section 2.1.3, embedding a lengthy target code string in a program's main function is not feasible.

For more complicated programs, the attack can be implemented separately in each function within a program. The additional setup required for expanding the attack to multiple functions is to make sure that the target program and the adversarial example program have the same number of functions and that those functions have the same names. This attacked source code format is illustrated in Figure 4.

This constraint on the functions ensures that the target program analysis aligns with the structure of the adversarial example program since the attack targets function analysis but not program structure. For example, if the target code analysis mentions calling function_foo, then the target LLM could produce unpredictable results or expose the attack if no function_foo is found. Similarly, if the target code does not cover a function_bar that is found in the adversarial example program, the LLM's analysis will not be consistent.

It is possible to include a single adversarial string in the main function that terminates the analysis in such a way that the LLM will terminate its analysis and not examine other functions. However, this method is more likely to be circumvented if the LLM decides to analyze functions besides the main function. It is also likely to be discovered through cursory human analysis if the executable binary has a large file size, but returns a program analysis that is too simple.

```c
#include <stdio.h>
#include <string.h>

// Function calculates sum, but is named getTemperature to align
// with target program
int getTemperature(void)
{
  int a = 4;
  int b = 9;
  int sum = 0;

  const char* command_string = "/* Prompt Injection Attack String */";
  const char* null_string = "";
  if (strlen(command_string) >= strlen(null_string)) {
    sum = a+b;
  }

  return sum;
}

int main(void)
{
  int result;
  result = getTemperature();

  if (result > 10) {
    printf("The sum is greater than 10.\n");
  }
  else {
    printf("The sum is 10 or less.\n");
  }

  const char* command_string = "/* Prompt Injection Attack String */";
  const char* null_string = "";
  if (strlen(command_string) >= strlen(null_string)) {
    printf("Calculated sum: %d\n", result);
  }

  return 0;
}
```

**Figure 4:** C code for an adversarial example with a main function and one other function.

The AutoDAN algorithm is not configured to find multiple successful prompt injection attack strings within a single input string. Instead of running these multi-function programs through AutoDAN, the successful prompt injection attack strings found in the single function examples were used in both functions of the multi-function source code files.

In the code example in Figure 4, the base program's function adds two hard-coded numbers then prints out if the sum of those numbers is greater than or less than 10. The target program's function is to compare two hard-coded values representing temperature values and then print a statement about the weather being cold or warm. The compiled modified source code performs the functions of the base program, but the automated analysis using Qwen3-8B determines the function to be that of the target program.

A second multi-function program was also tested. In the second example, the base program computes the factorial value of a number while the target program averages the values contained in an array. When the modified binary file is compiled and executed, it performs the function of the base program and the analysis again claims its behavior is that of the target program.

Both of these multi-function adversarial examples were also submitted for analysis using GPT-OSS-120B. The attack was also successful with analysis by this larger model. As in the single function analysis, GPT-OSS-120B discovered printed strings in the base programs. In one of the files, the automated analysis attributed the presence of this code to re-use of older code rather than as an indication of anything nefarious or deceptive.

## 4. Conclusion and Future Work

The introduction of automated analysis tools can greatly expand the workload capabilities of a static software analyst, but those same tools also introduce new areas of vulnerability. Attacking the LLM behind these AI-assisted tools undermines this enhanced capacity by requiring a human analyst to examine the code to determine its true functionality. The result is either code that is not properly identified or a loss of the efficiency benefit provided by the automation process.

To combat this vulnerability further work should be conducted to discover the full extent of the effectiveness of this attack, specifically how broadly it can be generalized to various LLMs including state-of-the-art commercial level models as well as how effectively it can be scaled to much more complex and potentially malicious programs. The executable binaries in the present research are simple programs; the results of implementing this same style of attack on more complex files needs to be examined further to determine broad applicability of these results. Future work should also be done to develop defenses to mitigate the indirect prompt injection attacks investigated in this research.

## Ethics Declaration

The research did not involve human subjects or data containing personally identifiable information (PII). Only computer systems administered by the authors were attacked for this research.

## AI Declaration

AI was not used in the drafting of this document. AI, however, was used in the development of this research for writing sample programs to test the method and for generating the seed text that the AutoDAN algorithm uses for generating prompt injection strings.

## Disclaimer

The opinions and views expressed are those of the authors alone and do not necessarily represent those of the U.S. Government, U.S. Department of Defense or its components, to include the Department of the Navy or the Naval Postgraduate School.